\documentclass{easychair}

\usepackage{newtxtext,newtxmath}
\usepackage{float}
\usepackage{biblatex} 
\usepackage{csquotes}
\usepackage[document]{ragged2e}
\nocite{*}

\title{HTML papers on arXiv: why it’s important, and how we made it happen}

\author{
Charles Frankston\inst{1}
\and
    A. Jonathan R. Godfrey\inst{2}
\and
    Shamsi Brinn\inst{1}
\and
    Alison Hofer\inst{1}
\and
   Mark Nazzaro\inst{1}}

\institute{
  arXiv\\
  \email{cbf@arxiv.org, shamsi@arxiv.org, ahofer@arxiv.org, mark@arxiv.org}
\and
   Massey University, New Zealand\\
   \email{a.j.godfrey@massey.ac.nz}
 }

\authorrunning{Frankston, Godfrey, Brinn, Hofer, Nazzaro}

\titlerunning{HTML papers on arXiv}

\begin{document}

\maketitle

\section{Introduction}
\label{sect:introduction}

arXiv is the world's largest and oldest scientific preprint server, and a champion of open science. Started in 1991, arXiv presently holds more than 2.4 million articles and is growing at an ever-increasing rate.

\begin{figure}[H]
	\begin{centering}
	\includegraphics[width=0.8\textwidth]{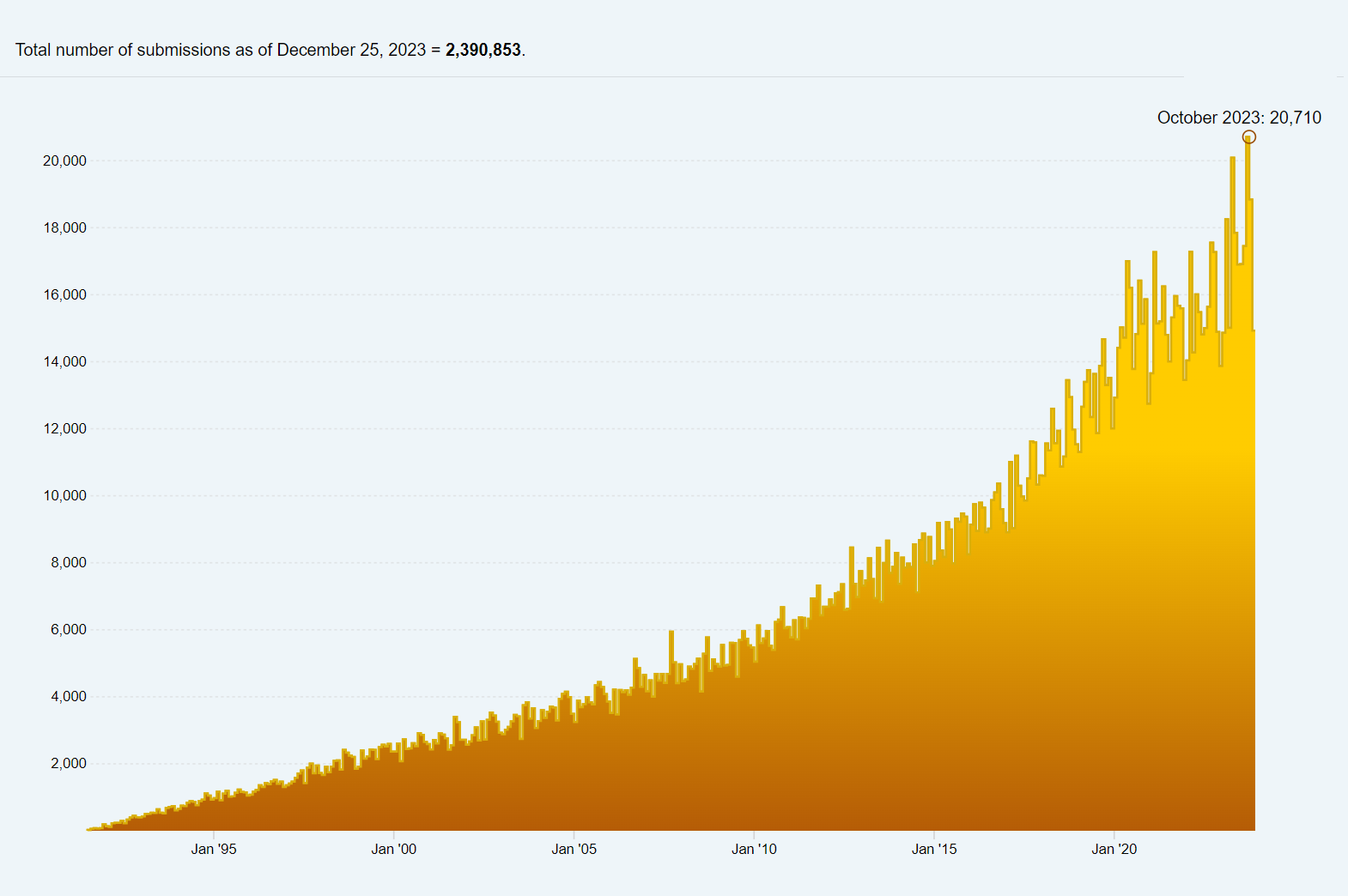}
	\caption{arXiv Monthly submissions from 1991 to present}
	\label{fig:monthly-submissions}
	\end{centering}
\end{figure}

In many fields of Physics, Math, and Computer Science, cutting edge research is first made available on arXiv. Examples:
\begin{itemize}
    \item LLM research (OpenAI, Deepmind, etc.)
    \item LIGO (Gravity wave research; 2017 Nobel Prize in physics)
    \item Proofs of famous theorems (Grisha Perelman)
\end{itemize}

arXiv has a mandate to continuously improve access to scientific research, and our long-term mission is to simply serve the needs of the scientific community through openness, collaboration and scholarship. Everyone has the right to participate in the wealth of scientific knowledge contributed to arXiv by researchers from all over the world. Accessibility is inherent to our mission of championing open science, and when we asked scientists with disabilities how arXiv could help make research more accessible they told us in no uncertain terms: add HTML as a format for papers.

Over the past few years, arXiv has made good progress in making \href{https://arxiv.org}{our website} more accessible according to W3C WAI guidelines. While this allows people with disabilities to more easily find and access papers, they often cannot read them because they are available almost exclusively in PDF format which has low native accessibility. 

What we heard from scientists with disabilities, standards experts, and accessibility researchers is that, when it comes to accessibility, PDF will always be playing catch up with HTML. Though we will retain PDFs on arXiv as always, adding HTML as a format will bring us closer to fulfilling the promise of truly open science.

\section{PDF limitations}
\label{sect:PDF limitations}

\subsection{Layout}
\label{sect:Layout}

PDF was originally designed as a digital representation of a printed page and there is frequently a mismatch between the paper’s geometry and the screen being used to view it. Zooming in to enlarge the text will often require horizontal scrolling. Responsive Design and content re-flow are not natural characteristics of PDF. Adobe is attempting to address these issues with features like “Liquid Mode”, but Liquid Mode is presently mobile-only and Adobe proprietary. It is not clear if this retroactive reinterpretation of PDF document layout will work well in all cases.

\subsection{Structure}
\label{sect:Structure}

Screen readers rely on structural elements with semantic meaning to efficiently map and navigate content. In a PDF, the original structure of the paper that screen readers depend on is typically lost. Section headings, captions, links, and more are reduced to typographic elements like font changes and positioning instructions. Mathematical notation in PDF is reduced to symbols and positioning directives, or simply turned into an image.

Adobe introduced tagged PDF to improve this situation, but despite its introduction in the early 2000s the vast majority of PDF documents in the world are not fully accessible. Properly tagging a PDF to make it accessible takes specialized knowledge and proprietary software. Presently, arXiv’s pipeline from \TeX~$\rightarrow$~PDF has poor support for tagged PDF (though the core \LaTeX\ team is working on a solution). Tagged PDF also has a sub-optimal solution for math: displaying it as an image with an alt-text description for screen readers.
\section{HTML}
\subsection{A better solution}
\label{sect:A better solution}

A lot of work has been done to make HTML accessible. When formatted correctly, HTML preserves the structure and intent of the document and renders fluidly on mobile devices. There is a rich ecosystem of assistive technologies and tools that build on this foundation. 

For most people, HTML is the gateway to all the content the internet has to offer. But PDF still dominates scientific publishing and there is a substantial body of scientific work online that is available only in PDF format.

Converting that PDF content to HTML poses challenges, even using modern AI-assisted techniques. The Allen Institute for AI created such a converter, SciA11y. While it produces some nice looking HTML documents, the job it is trying to do---reconstruct the paper's structure that was lost in the process of producing the PDF---is fundamentally difficult, and the SciA11y convert has serious limitations. To quote from \Textcite{wang2021improving}:
\begin{displayquote}
In the current iteration of the HTML render, we do not display author affiliations, footnotes, or mathematical equations due to the difficulty of extracting these pieces of information from the PDF.
\end{displayquote}
In conversations with the SciA11y team, they agreed that, in the case where source TeX files are available for a paper, it is better to produce HTML starting from that source than use their tool to convert from the PDF.

\subsection{A Strong HTML Ecosystem}
\label{sect:A strong HTML ecosystem}

Because use of the World Wide Web has become fundamental to human existence, a great deal of effort has been expended to make the web accessible. There’s a vibrant ecosystem of screen readers for HTML web pages and the open-source structure of HTML enables the creation of add-ins that can change page appearance to aid visually impaired and people with dyslexia. Dark mode is now widely supported for those with light sensitivities, while adjusting font size is trivial on a well formatted HTML page. Modern web browsers also build in useful features like language translation, which expands access in a different way (something that is at-present not smoothly supported in any PDF viewer we know of). An well-formatted article will re-flow to present nicely on small devices like mobile phones, which sometimes might be the only internet device available to scientists in poorer countries.

In the early days of the web, text in PDF documents was not indexed by search engines. This isn't the case anymore, but PDF documents make search engines work harder -- particularly PDFs where much of the text is actually represented in image form. By comparison, text harvesting from HTML documents is relatively easy. arXiv’s corpus is a substantial part of nearly all the Large Language Models today, and we've been told by researchers attempting to use our corpus that having HTML versions of documents available would support their work.

\section{\TeX\ and \LaTeX}
\label{sect:TeX and LaTeX}

TeX is a computer program that Donald Knuth wrote in the late 1970s to automate the typesetting of his books on the Art of Computer Programming~\cite{ieee2018tex-history1,ieee2019tex-history2}. It was born out of his frustration with the typesetting systems of the times, particularly as it related to mathematics and tables. \TeX\ is similar to other ``text formatters'' that existed at the time, such as runoff and Scribe, but is uniquely capable in that \TeX\ allows programming constructs that are powerful enough to make it Turing complete. This is both a blessing and a curse. The blessing is that users can create very sophisticated \TeX\ documents that layout their papers exactly as they wish. The curse is that it makes it difficult to re-interpret those papers for a target other than the printed page they were intended for.

\subsection{\LaTeX and reinterpretation}
\label{sect:LaTeX and reinterpretation}

\TeX’s powerful programming features enabled Leslie Lamport to create a package of extensions called \LaTeX~\cite{mittelbach23:latex}. \LaTeX\ defines tags/macros to create structured documents. Tags look like: \verb |\section| or \verb |\paragraph|. Since almost all scientific documents built with TeX use LaTeX now, in theory it should be easy to convert \LaTeX\ tags to HTML, for example: \verb |<h1>| or \verb |<p>|. In practice most \LaTeX\ papers have additional complexity. But the semantics of tags like \verb |\section| only partially match up to \verb |<h1>|. (\verb |\section| has more semantics than \verb |<h1>| --- mostly related to print formatting.) 

Very few papers submitted to arXiv use `base' \LaTeX.  Most commonly, the paper will use a package of settings and extensions intended to produce papers that are formatted in a particular manner.  Many of these packages are document templates provided by journals or societies so authors can deliver papers formatted according to that publisher's standards.

For example, the American Physical Society offers the Physical Review Journal's REVTeX template, built on \LaTeX. A document formatted using REVTeX will be ``publication ready'' for inclusion in an APS journal. These templates ease the burden for authors, but at the same time complicate the job of converting from \LaTeX. There are hundreds, if not thousands, of such templates; some are provided by large societies, some by small; some are just extensions shared amongst individual groups of scientists; and some eager scientists create extensions just for their own use, sometimes because they want to fine-tune the layout of their papers on a Letter size or A4 page. They might do this fine-tuning in ways that look quite wrong in any other medium or format.

\section{So what’s the problem?}
\label{sect:So whats the problem?}

90\% of arXiv's papers are submitted as TeX (lately almost all LaTeX) format. So why can't we just run the TeX compiler to output to HTML instead of PDF?

Unfortunately that is not supported and is not easy to support. Donald Knuth was concerned with the precise typesetting of his technical content. The TeX compiler he released in 1978 is exclusively intended for that purpose. Even though LaTeX adds structure to the TeX language, that structure is almost completely lost at compile time when the LaTeX constructs are converted back to TeX primitives. 

The core LaTeX team (see \href{https://www.latex-project.org/latex3/}{latex-project.org}), headed by Frank Mittelbach, is working on revising \TeX/\LaTeX\ processing to preserve the structured information. They are also adding new constructs to \LaTeX\ for things like alt-text. Currently, their work is exclusively focused on producing better tagged PDF and is several years from completion. But we are hopeful that as this project nears completion it will be easy to produce structured HTML with a small amount of additional work.

\subsection{But my journal offers HTML now}
\label{sect:But my journal offers HTML now}

Many journals offer HTML for the online versions of their papers, including those submitted as \LaTeX. Why can't arXiv use the same techniques that journals use? The simple answer is that the journals accomplish this through a combination of tooling (that might even be specific to that journal's \LaTeX\ template), and manual labor. There are a specialized providers of formatting and accessibility remediation for publishers that can perform this work as a ``reasonable'' cost. But while it may be reasonable for a journal to spend \$30, \$50, or even \$100 to prepare an article for publication, it's not feasible for arXiv. We now process around 20,000 papers per month, with most papers getting announced within 24 hours of submission. There is simply no budget for manual or custom processing, and we have not found any service that can keep up with our announcement pace.

It must continue to be the case that the labor for producing an article on arXiv is provided entirely by the author. At the same time, we cannot burden our authors with additional work to generate an HTML version of their papers, or to become accessibility experts. (Note: we are not averse to encouraging authors to make small adjustments to their submissions to help us produce better HTML, but it should be reasonably easy and because the authors wish to do so.) So we began our quest for an entirely automated solution. 

\subsection{arXiv’s pragmatic approach}
\label{sect:arXivs pragmatic approach}

We know that researchers with disabilities need solutions now and did not want to wait for the \LaTeX\ team to finish their tagging work, so we investigated existing tools. There have been at least ten attempts to create \TeX/\LaTeX\ to HTML tools (see \href{https://texfaq.org/FAQ-LaTeX2HTML}{texfaq.org} for a partial list).

We evaluated all the tools we could find and the conclusion was that the best tools were:

\begin{itemize}
\item LaTeXML maintained by Bruce Miller and Deyan Ginev at NIST (National Institute of Standards and Technology)
\item Tex4ht, created by Eitan M. Gurari, now maintained by Michal Hoftich
\end{itemize}

These two tools were roughly tied in the quality of the HTML produced, but LaTeXML has a larger library of supported packages, and the predecessor ar5iv Labs project at arXiv used LaTeXML, which made it a logical choice.

\subsection{The ar5iv Project}
\label{sect:The ar5iv project}

\href{https://ar5iv.labs.arxiv.org}{ar5iv} was a project started by Dr. Michael Kohlhase from KWARC, and Ph.D. student Deyan Ginev (see \cite{stamerjohanns2010transforming}). The intent was to offer HTML versions for arXiv’s entire LaTeX corpus, using LaTeXML for the conversions. 

\begin{figure}[H]
    \centering
    \includegraphics[width=1.0\linewidth]{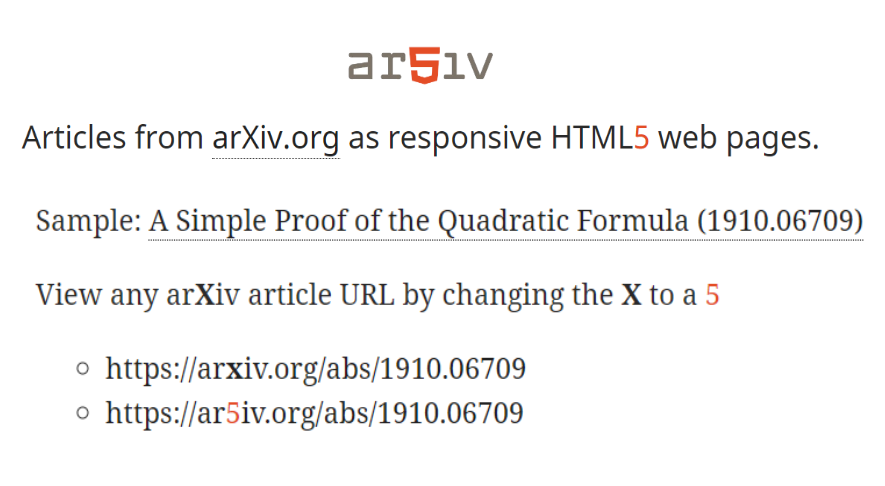}
    \caption{ar5iv.org home page}
    \label{fig:enter-label}
\end{figure}

It mostly succeeded at this, but some 25\% of papers had conversion errors. (Note that papers with conversion errors may still be readable. Only 3\% totally fail conversion). As LaTeXML coverage has improved success rates have increased and errors decreased, but old articles must be reconverted to remove glitches. This is computationally expensive. At an average of \$0.015 per article on Google Cloud, it would cost \$30,000 every time we wanted to reconvert the entire arXiv corpus. Fortunately, the KWARC team in Germany has generously provided the ar5iv project with access to compute resources to run the conversions, but it takes several weeks to re-convert the entire corpus.

In many ways, arXiv's current effort to offer HTML for all papers, in parallel to the PDF, can be considered bringing the HTML effort in-house.

\subsection{The difficulty of converting \LaTeX to HTML}
\label{sect:The difficulty of converting LaTeX to HTML}

LaTeXML, like nearly all the TeX to HTML tools, does not use the actual TeX compiler. As explained earlier, that compiler in its present state loses structural information. LaTeXML therefore does its own processing of the \TeX/\LaTeX\ documents and packages, emulating just enough of TeX to preserve the information provided by the author. Since the LaTeXML compiler is does not (yet) implement a complete \TeX\ engine many macros and extensions have to be handled as special cases. That is, the LaTeXML converter has built in code to handle the use of some packages, rather than trying to interpret the actual package definition as the core \TeX\ engine would. The LaTeXML team has done such special case implementations for more than 400 of the most commonly used \LaTeX\ packages in arXiv, but that still leaves a considerable number of packages it does not recognize.

\subsection{Unfinished work}
\label{sect:Unfinished work}

In addition to ``one-offs''---papers where clever scientists defined their own macros and extensions---there is still a long tail of less common packages, and some popular packages that are hard to implement are not yet supported as well. The most substantial example is \href{https://tikz.dev/}{tikz}, a popular diagramming package that produces vector graphics from a geometric description.

The LaTeXML team is currently working on improving their raw interpretation of TeX to automatically cover the long-tail of missing packages, reducing the number that need to be handled as special cases. And there are still some \LaTeX\ constructs that in edge cases don’t render quite right. The present conversion results are mostly successful, but many papers still have rendering glitches, although most are minor. Offering HTML papers now, even with rendering glitches, is far better than nothing and brings immediate benefits to scientists with disabilities and improves legibility on mobile devices.

Even papers that might use an unsupported package can still be useful for readers, as the paper might only use the construct defined in that package in one or two places. If the LaTeXML converter doesn't recognize a \LaTeX\ command (because it didn't know about the package that defined it), it will simply output that command to the text of the paper. arXiv readers who are working members of the scientific community will often be able to deduce the meaning of the construct, and thus still be able to fully understand the content of the paper. This would also be true for those using screen-readers who have some familiarly with \LaTeX.

\subsubsection{Sample HTML Paper}
\label{sect:Sample HTML paper}

\begin{figure}[t!]
    \centering
    \includegraphics[width=0.8\linewidth]{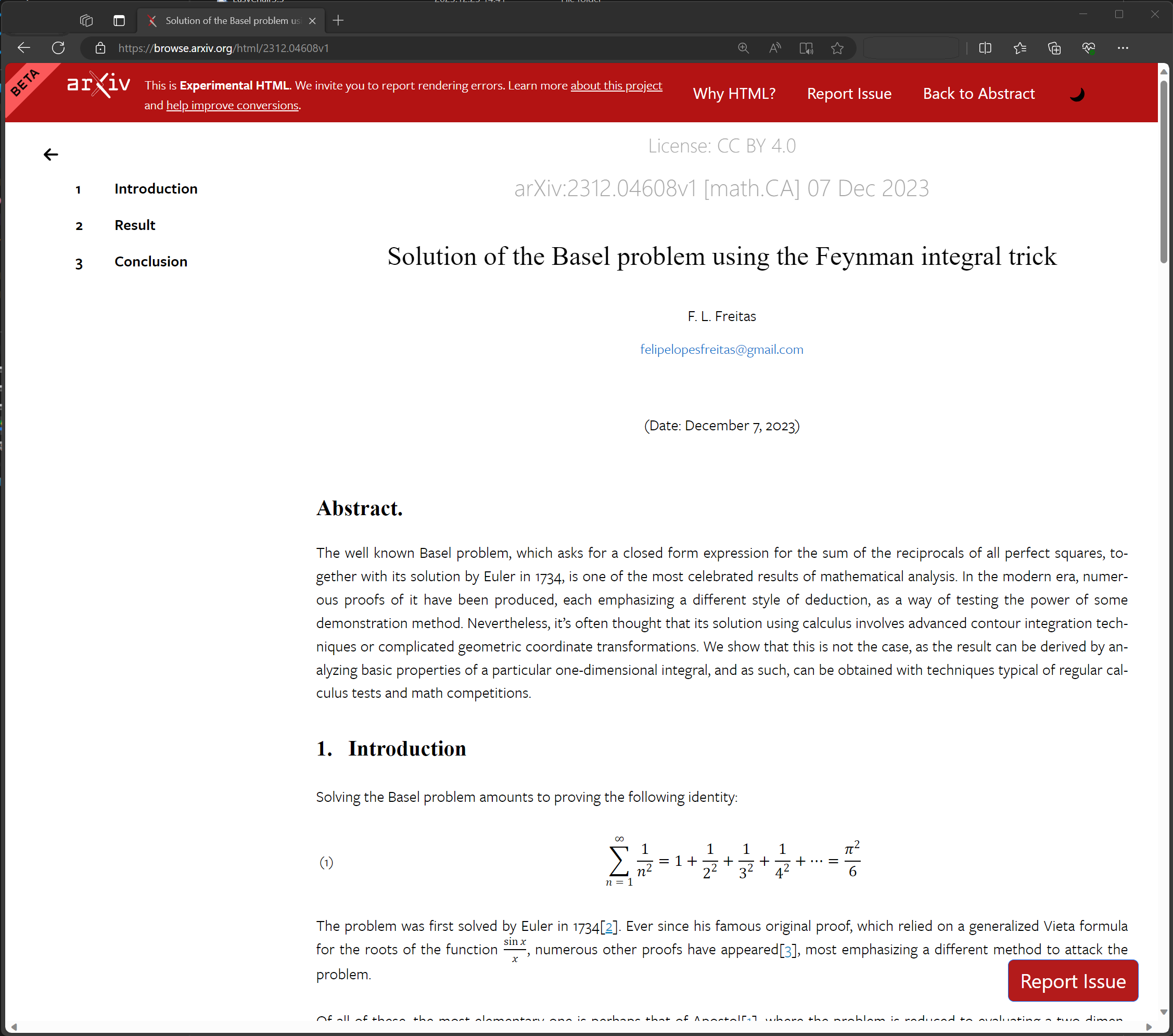}
    \caption{Sample HTML paper}
    \label{fig:sample-html}
\end{figure}

\begin{figure}[b!]
    \centering
    \includegraphics[width=0.8\linewidth]{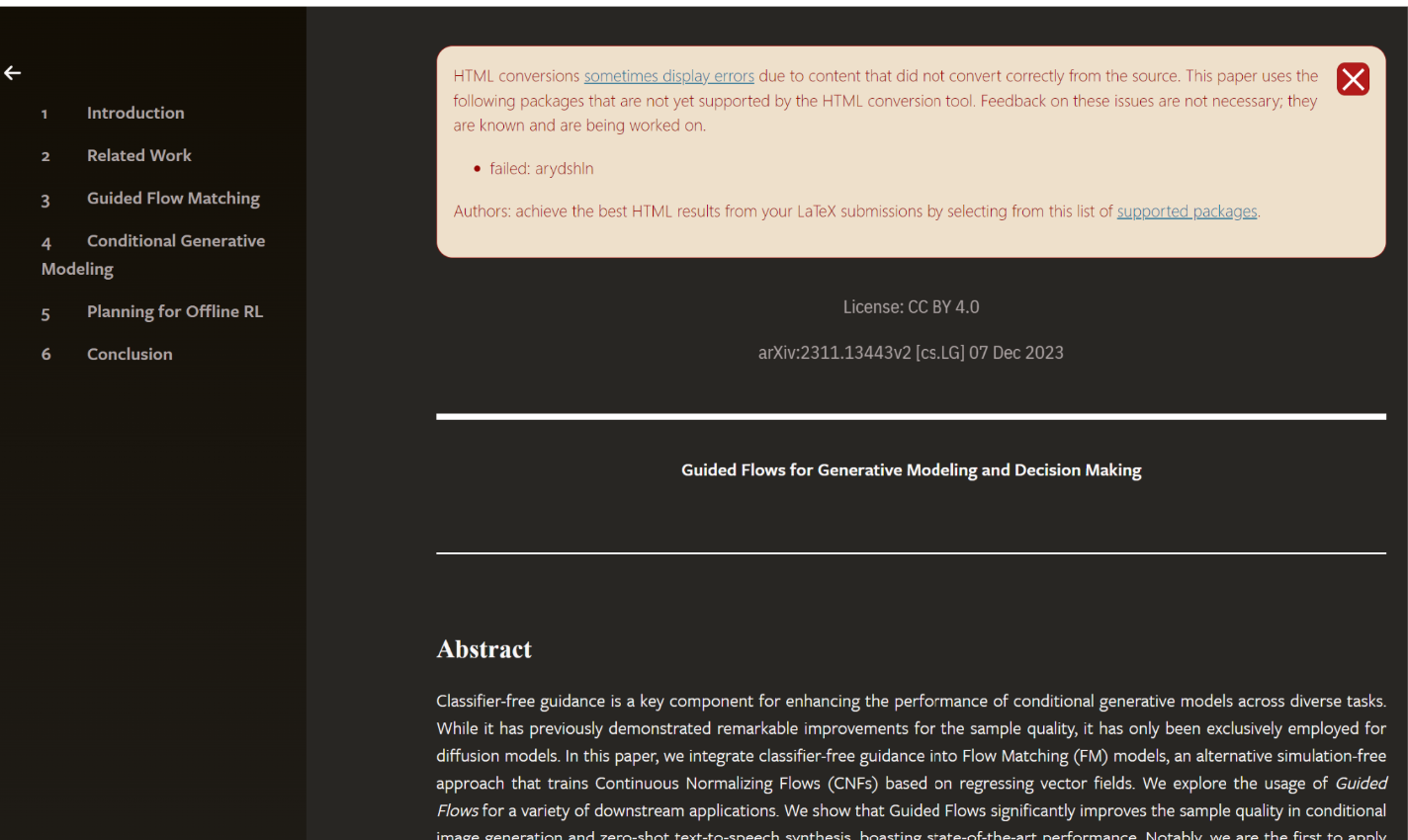}
    \caption{HTML paper sample, dark mode}
    \label{fig:dark-mode}
\end{figure}

Figure \ref{fig:sample-html} below shows a sample HTML paper. Note:
\begin{itemize}
    \item the ``Experimental'' label
    \item the prominent ``Report Issue'' button
\end{itemize}

We are describing the whole HTML effort as an experiment. This is to make the arXiv community aware that the present iteration is not perfect, and to expect improvements in the future as the ``experiment'' continues.

The ``Report Issue'' button is present on all the HTML documents so the arXiv community can help us identify issues in LaTeXML. The developers of LaTeXML indicated to us that they would welcome bug reports on LaTeX constructs that did not render or read correctly. The button automatically records which paper the issue is on and captures a snippet of the relevant content when there is an active text selection. This has quickly produced a number of bug reports, largely of good quality. The main issue now is duplicate issue reports. We plan to improve the bug reporting feature and help reduce duplicates by showing users the reports already received on the same article.

Figure \ref{fig:dark-mode} above shows a paper rendered in dark mode, and a message shown at the top of any paper that called for package that LaTeXML didn't know about -- an indication to readers that most likely something in this paper isn't going to render correctly. Dark mode is automatically chosen if that is the current setting for the user's browser. Dark mode has become a popular preference in recent years and is a common accessibility issue for those with light sensitivities or viewing their screen under sub-optimal conditions like glare or low light.

\section{Early Impact}
\label{sect:Early Layout}

We rolled out HTML formatted papers on arXiv on December 1, 2023. These are early days, but the reaction from the scientific community has been very positive. Most importantly, initial testing with screen reader and Braille users confirm that it is a much better experience than PDF. One early tester told us ``pleasingly, in the most part the Braille translation is very good here.'' We heard from another that ``the reading of the HTML file by VoiceOver is impressive even for mathematical formulae.'' We have also heard from many people who can now read papers on their mobile devices, and appreciation for being able to easily adjust font size and use dark mode. 

This early feedback is heartening, but there is a long way to go. As noted earlier, we are receiving feedback from authors and readers about errors and issues they are coming across in some papers. However, we are keeping in mind that scientists with disabilities asked us to not let the perfect be the enemy of the good. Our ongoing challenge is to balance author's desires to present the highest quality version of their work with the tremendous accessibility gains from even imperfect HTML. We look forward to building on this important foundation in the future weeks, months, and years. 

\section{Future Work}
\label{sect:Future Work}

\begin{itemize}
    \item Continue to work with the LaTeXML team to improve the conversion process.
    \item Figure out a cost-effective way to periodically re-compile the whole corpus to pick up these improvements.
    \item Revisit tooling in a few years when the \LaTeX\ team is further along. We are hopeful that the work to produce tagged PDFs will also enable the generation of HTML output using the core \TeX\ engine.
    \item Make charts and graphs more accessible, possibly by providing a way for users to access the data behind the graph.
    \item Explore pathways to better image captions, including auto-caption or crowd-sourcing missing alt-text. It is a pressing accessibility issue and challenging AI problem. An early user tester shared a long-standing wish that ``it would be great if there was a system that incentivised authors to provide their own alt text.''
    \item Continue to listen to feedback from authors, and share best practices as we learn more, working together to improve HTML papers on arXiv.
\end{itemize}

\section{Acknowledgements}
\label{sect:Acknowledgements}

Kudos to Paul Ginsparg \& co for insisting from the start in 1991 that scientists submit the source code for their papers, instead of just the PDF, which means we can now at least try to convert the 90\% of our corpus that is in \TeX\ to HTML.

We thank the many scientists with disabilities who so generously shared their expertise, insights, and feedback, and guided arXiv's efforts in the most useful direction.

This material is based upon work supported by the National Science Foundation under Award No. OAC-2311521 and by NASA under award No. 20-OSTFL20-0053. Any opinions, findings and conclusions or recommendations expressed in this material are those of the author(s) and do not necessarily reflect the views of the National Science Foundation or of NASA.

\printbibliography 


\end{document}